\documentclass{aa} 
\usepackage{txfonts}
\usepackage{multirow}
\usepackage{graphicx,epsf}
\usepackage{url}
\usepackage{amsmath,graphicx,adjustbox,setspace, sidecap, float}
\usepackage[toc,page]{appendix}
\usepackage{gensymb}
\usepackage{fancyhdr}
\usepackage{natbib}
\usepackage{caption,subcaption}
\usepackage{threeparttable}
\usepackage{color}

\usepackage[dvipsnames]{xcolor}

\newcommand{\sig}{\Sigma}

\newcommand{\Me}{M_{\oplus}}
\newcommand{\Mj}{M_{\rm J}}
\newcommand{\Msun}{M_{\odot}}
\newcommand{\OmegaK}{\Omega_{\rm K}}

\newcommand{\Tb}[1]{Table~\ref{tab:#1}}

\usepackage[backref, pagebackref=false, hyperindex=true, breaklinks=true, colorlinks=true, urlcolor=magenta, linkcolor=magenta, citecolor=NavyBlue, citebordercolor={0 1 0}, pagecolor=red, bookmarks=true, filecolor=blue,bookmarksopen=true, plainpages=false, pdfpagemode=UseThumbs]{hyperref}

\begin{document}

\title{Hydrodynamical Simulations of Resonant Breaking in Multi-Planet Systems via Rebound Migration During Disk Dispersal}

\author{Beibei Liu\inst{1,2 \star},  Cl{\'e}ment Baruteau\inst{3}, Zhaohuan Zhu\inst{4}, Ya-Ping Li\inst{5}, Sijme-Jan Paardekooper \inst{6} \and Shigeru Ida\inst{7}  }

 \institute{Institute for Astronomy, School of Physics, Zhejiang University, Hangzhou 310027,  China \ 
 \ \email{bbliu@zju.edu.cn}
\and
  Center for Cosmology and Computational Astrophysics, Institute for Advanced Study in Physics, Zhejiang University, Hangzhou 310027,  China
\and
IRAP, Universit{\'e} de Toulouse, CNRS, CNES, Toulouse, France 
\and
Department of Physics and Astronomy, University of Nevada, Las Vegas, NV 89154-4002, USA
\and           
Shanghai Astronomical Observatory, Chinese Academy of Sciences, Shanghai 200030, China  
\and 
TU Delft, Faculty of Aerospace Engineering, Kluyverweg 1, 2629 HS Delft, The Netherlands
\and 
Department of Astronomy, School of Science, Westlake University, Hangzhou, Zhejiang 310030, China
}

  \abstract
   {
This study extends the investigation of rebound outward migration to multi-planet systems near an inner expanding disk cavity driven by stellar X-ray photoevaporation. Using 2D hydrodynamical simulations, we explore how systems of two and three planets that span masses from super-Earths to Jupiters evolve as the disk disperses from the inside out. Our results show that rebound migration can substantially reshape multi-planet architectures in the final stages of disk clearing. Owing to the strong, positive corotation torque exerted onto the planet near the cavity edge, divergent migration of the neighbouring planets can break resonant configurations and trigger dynamical instabilities, producing non-resonant orbits with widened period ratios. However, the outcome depends critically on planet mass and the disk dispersal timescale. In lower-mass disks where cavity expansion is too rapid, rebound migration is suppressed, and systems tend to preserve resonant chains. These findings suggest that the rebound mechanism can provide a compelling pathway to explain the prevalence of widely separated, non-resonant architecture observed in the exoplanet population.
   }
    
\titlerunning{Rebound migration in photo-evaporating disks}
\authorrunning{Beibei Liu et al.}
\maketitle

\section{Introduction} 
\label{sec:introduction}

Planet orbital migration driven by planet--disk interactions is a fundamental process shaping planetary system architectures. While migration in smooth disks has been extensively studied \citep{Kley2012,Baruteau2013,Baruteau2014,Paardekooper2023}, the dynamics near disk boundaries, remaining far less explored, represent a distinct and crucial regime \citep{Masset2006,Ogihara2010,Liu2017a,Brasser2018,Liu2025}.

When the gas surface density gradient at a cavity edge becomes sufficiently steep, the corotation torque can become strongly positive to dominate over the Lindblad torque, reversing the direction of inward migration \citep{Masset2006,Liu2017a,Romanova2025}. This mechanism was identified in the context of magnetospheric cavities and has been termed ``rebound'' migration \citep{Liu2017a}. Recent work has extended this concept to photoevaporative cavities, which open at larger radii (typically a few~au) and expand outward as the disk disperses \citep{Liu2022}. Hydrodynamical simulations have verified that a single planet near such an expanding edge can be pushed outward, with the rebound rate depending sensitively on planet mass and disk conditions \citep[hereafter Paper~I]{Liu2025}.

In multi-planet systems, the picture becomes more complex due to mutual gravitational interactions. Convergent migration often leads systems to capture into mean-motion resonances, but analytical work of \cite{Liu2017a} proposed that an expanding cavity edge can drive resonant pairs outward and potentially pull them apart. The observed prevalence of near- or non-resonant planet pairs in transit and radial velocity surveys \citep{Fabrycky2014,Dai2024,Hu2025} suggest the need for mechanisms that alter the resonant configurations during/after the late-stage gas disk evolution 
\citep{Liu2017b,Pichierri2024,Pan2025,Hansen2025}; also see \citealt{Papaloizou2011,Baruteau2013,Izidoro2017,WuYQ2024,Lin2025,Lin2026}.

Paper I provided the first hydrodynamic demonstration that a single planet near an expanding photoevaporative cavity edge can undergo outward rebound migration. In this work, we extend such investigations to multi-planet systems at the same type of disk edge.  
Using high-resolution simulations, we study how planet pairs with varying mass ratios and initial orbital configurations evolve under the combined effects of disk torques and mutual interactions.  We examine whether such planet pairs remain in or escape from the resonant configurations. Our results can provide key inputs for future dedicated N-body studies aiming to explain the orbital architectures of observed exoplanet systems.

The paper is structured as follows. In Section~\ref{sec:method}, we introduce our numerical setup and model parameters. Sections~\ref{sec:pair} and \ref{sec:multi}  describe the migration outcomes for two-planet pairs and multi-planetary systems with various masses and initial orbital configurations. 
We present the discussions and conclusions in Section~\ref{sec:discussion} and Section~\ref{sec:conclusion},  respectively.

\section{Method}
\label{sec:method}

\begin{table*}
    \centering
     \caption{Main parameters of the simulations}
    \label{tab:param}
    \begin{tabular}{lllcccc}
    \hline\hline 
   symbol & description  & values  \\ 
\hline
\centering   $\Sigma_0$ & gas surface density at 10 au      &  $44 ~ \rm g/cm^{2}$ \  \ (fiducial disk) \\
\centering & & $13  ~ \rm g/cm^{2}$ \ \ (less massive disk)  \\
\centering   $s$ & gradient of gas surface density, ${=}d\log\Sigma / d\log r$      &  -0.5  \\
\centering  $h_0$ & disk aspect ratio at 10 au &    0.05  \\
\centering  $k$ & gradient of disk aspect ratio, ${=}d\log h / d\log r$    &  0.25  \\
\centering  $\alpha$ & turbulent viscous parameter     &  $10^{-3}$ \\
\centering  $\dot M_{\rm PE}$ & total integrated mass loss rate from \cite{Owen2012}     &  $5 \times 10^{-8} ~ \rm M_{\odot}/yr$ \\
\hline
    \end{tabular}
    \label{tab:IC}  
\end{table*}

\begin{table*}
    \centering
    \caption{List of simulations}
   \label{tab:planets}  
   \begin{threeparttable}  
    \begin{tabular}{lccccccc}
    \hline\hline 
 list & run$^{a}$  & planet mass & initial period ratio   & rebound$^{b}$ & final period ratio  \\ 
    & & $[M_{\rm k}, M_{\rm k+1}]$
     &  $P_{\rm k+1}/P_{\rm k}$ & \\
\hline
\centering  1 & \texttt{fid\_M02\_R1.5}  & $5~\Me, 20~\Me$ &  3:2   & Yes  & 2.85    \\ 
\centering 2 &  \texttt{fid\_M20\_R1.5}  & $20~\Me, 5~\Me$ &  3:2   & Yes  & 1.75    \\
\centering 3 &  \texttt{fid\_M02\_R2.0}  & $5~\Me, 20~\Me$ &  2:1   & Yes & 2.93    \\ 
\centering 4 &  \texttt{fid\_M20\_R2.0}  & $20~\Me, 5~\Me$ &  2:1   & Yes  & 2.46     \\
 \centering 5 &  \texttt{fid\_M21\_R1.5}  & $20~\Me, 10~\Me$ &  3:2   & Yes  &2.20     \\
  \centering  6 &  \texttt{fid\_M12\_R1.5} & $10~\Me, 20~\Me$ &  3:2   & Yes  & 2.20    \\
\centering 7 &  \texttt{fid\_M21\_R2.0}  & $20~\Me, 10~\Me$ &  2:1   & Yes  & 1.93    \\
 \centering 8 &  \texttt{fid\_M12\_R2.0}  & $10~\Me, 20~\Me$ &2:1   & Yes   & 2.31    \\
 \centering 9 &  \texttt{fid\_M23\_R1.5} & $20~\Me, 0.3~\Mj$ &  3:2  & No & 1.50     \\
 \centering 10 &  \texttt{fid\_M32\_R1.5}  & $0.3~\Mj, 20~\Me$ &  3:2   & Yes  & 3.73     \\
 \centering 11 &  \texttt{fid\_M23\_R2.0} & $20~\Me, 0.3~\Mj$ &  2:1  & No &  1.50   \\
 \centering  12 &  \texttt{fid\_M32\_R2.0}  & $0.3~\Mj, 20 ~\Me$ & 2:1   & Yes  & 3.93    \\
  \centering  13 &  \texttt{fid\_M34\_R1.5} & $0.3~\Mj, 1~\Mj$ & 3:2   & No &  1.50    \\
 \centering 14 &  \texttt{fid\_M43\_R1.5}  & $1~\Mj, 0.3~\Mj$ & 3:2   & Yes  & 6.41    \\
   \centering 15 &  \texttt{fid\_M34\_R2.0}  & $0.3~\Mj, 1~\Mj$ &  2:1   & No   & 2.02     \\
 \centering 16 &  \texttt{fid\_M43\_R2.0}  & $1~\Mj, 0.3~\Mj$ & 2:1  & Yes   & 6.47     \\
  \centering 17 &  \texttt{fid\_M123\_R1.5}  & $10~\Me, 20~\Me, 0.3~\Mj$ & 3:2, 3:2  & Yes   & [1.50, 3.98]     \\
    \centering 18 &  \texttt{fid\_M123\_R2.0}  & $10~\Me, 20~\Me, 0.3~\Mj$ & $\gtrsim$ 2:1, 2:1  & Yes  & [2.29, 3.11]   \\
   \centering 19 &   \texttt{lowdisk\_M123\_R1.5}  & $10~\Me, 20~\Me, 0.3~\Mj$ & $\gtrsim$ 3:2, 3:2  & No   & [1.50, 1.50]     \\
    \centering 20 &   \texttt{lowdisk\_M123\_R2.0}  & $10~\Me, 20~\Me, 0.3~\Mj$ & $\gtrsim$ 2:1, 2:1  & No   &   [1.50, 2.01]  \\
\hline
    \end{tabular}
    a. The number $0$-$4$ following M refers to planets of $5~\Me$, $10~\Me$, $20~\Me$, $0.3~\Mj$ and $1~\Mj$, respectively, while $1.5$ and $2.0$ following R refer to initial period ratio slightly away from the 3:2 or 2:1 mean-motion resonances. \\
    b. Yes refers to the magnitude of rebound migration of any planets being larger than $5\%$.
    \end{threeparttable}  
\end{table*}

Following Paper I, we use the $2$D grid-based hydrodynamical code \href{https://github.com/charango/dustyfargoadsg}{\texttt{Dusty FARGO-ADSG}} to investigate multi-planet migration in a photo-evaporating disk during its gas dispersal phase. Our simulations only model the gas content of the disk, and the disk self-gravity is discarded. 
The governing equations are expressed as 
\begin{equation}
\begin{aligned}
& \frac{\partial{\sig} }{\partial{t}} + \nabla   \cdot  \left( \sig \vec{v}  \right) = \dot \Sigma_{\rm PE}, \\
& \frac{\partial{\vec{v}} }{\partial{t}} + \vec{v} \cdot \nabla \vec{v} = -\frac{\nabla P}{\sig} - \nabla \Phi + \vec{f}_{\nu},
 \label{eq:conservation}
\end{aligned}
\end{equation}
where $\Sigma$, $P$, $\vec{v}$, and $\vec{f}_{\nu}$   represent the gas surface density, pressure, velocity and viscous force per unit mass, and $\Phi$  is the total gravitational potential, which includes the direct contributions from the star and the planet, as well as the indirect contribution from the planet due to the acceleration of the coordinate frame (see Eq. 2 of Paper I). A locally isothermal equation of state is assumed. We adopt the mass-loss rate profile from \cite{Owen2012}, where $\dot{\Sigma_{\rm PE}}$ is peaked in the inner disk region of ${\sim}2{-}3$ au.  
This code allows us to self-consistently model the co-evolution of the planets and disk under the influence of turbulent viscous accretion and stellar X-ray photo-evaporation (PE).  The code units for length, mass, and time are defined as $r_0{=}10$ au, $M_0{=}M_{\star}{=}1 \Msun$, and $ t_0 {=}  \Omega_{0}^{-1}{=} \sqrt{r_{0}^3/G M_0}$ (see \Tb{param}), where $\Omega_0$ is the Keplerian angular frequency at $r_0$ and $G$ is the gravitational constant. We also define $P_0 {=} 2\pi/\Omega_0$, ${\approx}31$ yr as the canonical time unit for figures.

The gas turbulent viscosity is parameterized using the Shakura-Sunyaev 
$\alpha$-prescription: $\nu {=} \alpha c_{\rm s} H$, where $\alpha$ is set to $10^{-3}$, $c_{\rm s}$ is the sound speed, $H {=} c_{\rm s}/\Omega_{\rm K}$ is the disk's pressure scale height, $h{=}H/r$ is the disk aspect ratio,  and $\OmegaK{=}\sqrt{G M_{\star}/r^3}$ is the Keplerian angular frequency at radial distance $r$. The initial disk surface density and aspect ratio are given by $\sig_0(r) {=} \sig_0 (r/r_0)^s$ and $h(r) {=} h_0 (r/r_0)^k$, with default parameters listed in \Tb{param}.

In this work we choose a set of fiducial parameters that differs slightly from that of Paper I in order to reduce computational cost. 
While the parameters in Paper I were already computationally expensive for single-planet simulations, our new configuration, featuring a photo-evaporation rate $1.5$ times higher and a slightly lower disk aspect ratio of $h_0 {= }0.05$, keeps computational demands modest enough for simulating the rebound migration of multiple planets throughout the disk dispersal phase.

Like in Paper I, grid cells are sampled logarithmically in radius between $0.2 \ r_0$ and $10 \ r_0$, and linearly in azimuth over the full $2\pi$. Two resolution grids are used: a standard resolution ($N_{\rm r}{=}480$, $N_{\phi}{=}770$) for most cases, and a high resolution ($N_{\rm r}{=}1200$, $N_{\phi}{=}1900$) specifically for the planet pairs with masses of $5 \ M_{\oplus}$ and $20 \ M_{\oplus}$. This choice  of $N_{\rm r}$  ensures that the full width of the planet’s horseshoe region (${\approx} 2.4r_{\rm p} \sqrt{q_{\rm p}/h}$, $q_{\rm p}$ is the planet-to-star mass ratio) is resolved by approximately ten radial cells, which is desirable for accurate corotation torque calculations. The corresponding $N_{\phi}$ values yield approximately square grid cells for better numerical accuracy.

To optimize computational efficiency, we first evolved the disk in $1$D ($N_{\phi}{=1}$) without planets until an inner cavity began to form, which is defined as the gas surface density dropping below $10^{-5} \rm \ g/cm^{2}$. We then restarted the simulations in $2$D, introducing planets at their full masses. The $1$D simulations used standard open boundary conditions at both radial edges, while the $2$D simulations employed wave-damping zones at these edges to prevent reflections of planet-induced wakes. We refer readers to Paper I for further numerical details.

We consider five planet masses: $5 ~\Me$, $10 ~\Me$, $20 ~\Me$, $100 ~\Me$, and $318 ~\Me$, representing
 super-Earths with two characteristic masses, Neptune-like planets, and gas giants similar to Saturn ($0.3 \ M_{\rm J}$) and Jupiter ($1 \ M_{\rm J}$). Various combinations of these five planet masses are adopted for the inner and outer planets.

 We consider the planets are initially placed close to the first order mean motion resonances, either the $3$:$2$ or $2$:$1$ resonances. These two states are most easily captured during long-range convergent disk migration \citep{Quillen2006,Kajtazi2023,Lin2025} and are also prevalent among observed multi‑planet systems \citep{Fabrycky2014}. The innermost planet begins at $r_0{=}10$ au in our $2$D simulations, while the outer planet's orbit is placed slightly exterior to the exact resonant location (e.g., at $16.5$ au for the $2$:$1$ resonance).

We also calculate the disk torque exerted onto the planet $\Gamma$, which is calculated as  
\begin{equation}
\begin{aligned}
 \Gamma =  \int_{r }  \int_{\phi} \Sigma \frac{ \partial \Phi_{\rm p}}{\partial \phi} rd r d \phi = \sum_r T(r),
\end{aligned}
\label{eq:Gamma}
\end{equation} 
where $\Phi_{\rm p}$ is the planet's gravitational potential, $T(r)$ is the torque density that represents the $\phi$-averaged torque across the radial grid cells.

\begin{figure*}
    \centering
    \includegraphics[width=0.96\hsize]{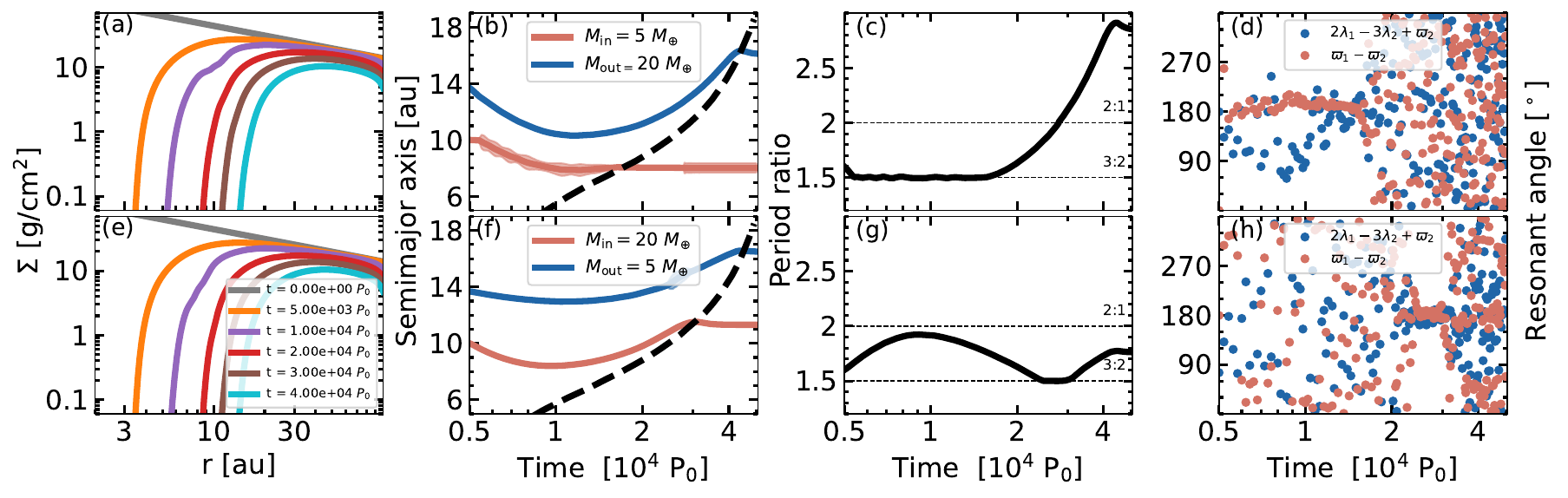}
    \caption{  
Evolution of disk surface density, planet semimajor axes, period ratios, and resonant angles for a two-planet system consisting of $5~\Me$ and $20~\Me$ planets. The top (bottom) row shows the configuration in which the inner planet is less (more) massive.
In the second column, the semimajor axes of the inner and outer planets are given by the red and blue curves, respectively, with shaded regions representing their perihelion and aphelion distances. The black dashed line marks the inner edge of the expanding cavity, defined as the radius where the surface density falls below $0.1~\rm g/cm^2$. The fourth column plots the resonant angles $2\lambda_1 - 3\lambda_2 + \varpi_2$ (blue) and $\varpi_1 - \varpi_2$ (red), where the subscripts $1$ and $2$ denote the inner and outer planets, respectively.
}
    \label{fig:semi_2pL}
\end{figure*}

\begin{figure*}
    \centering
    \includegraphics[width=0.85\hsize]{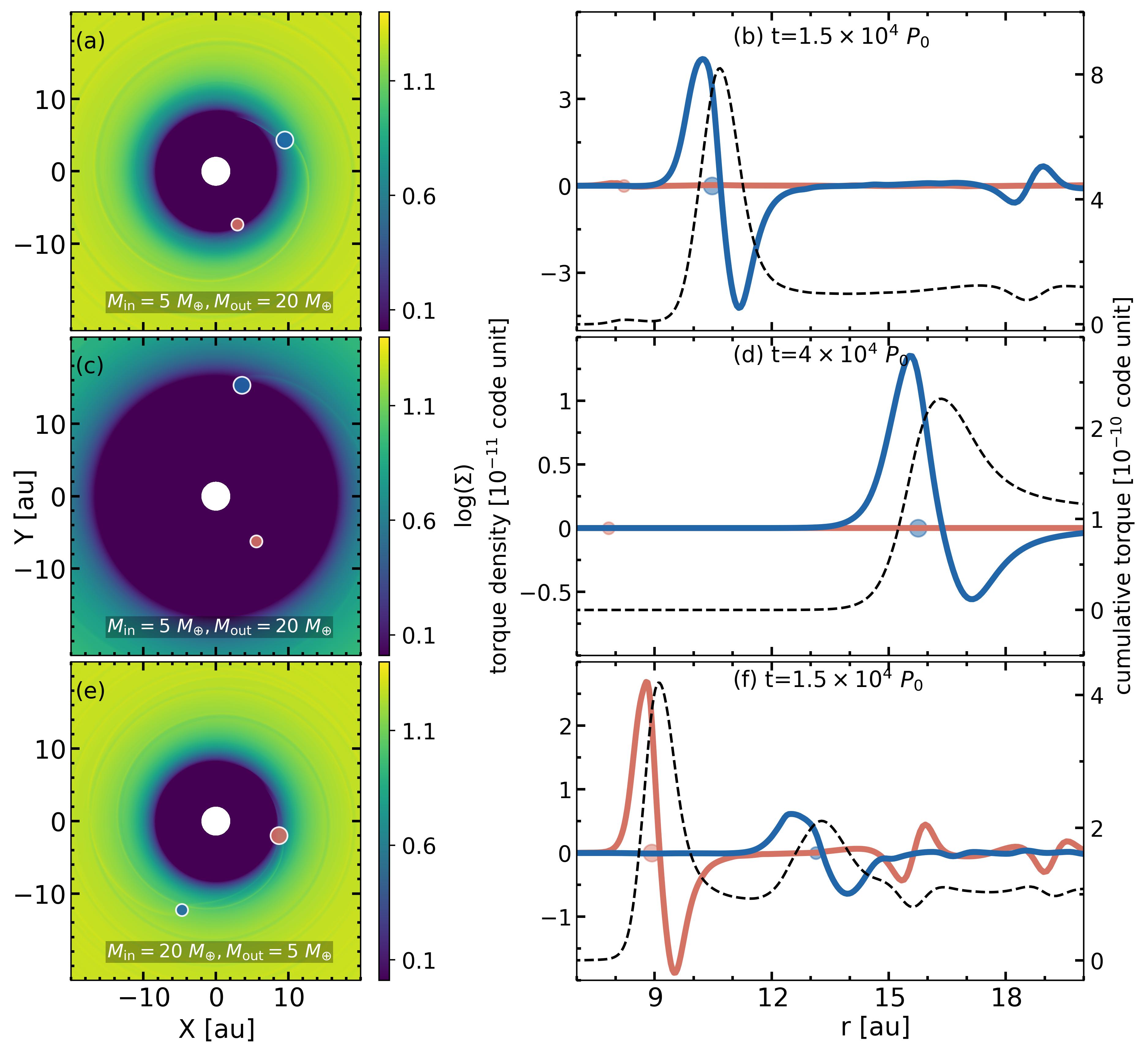}
    \caption{  
Snapshots from the data in Figure 1 of surface density $\rm \log(\Sigma)$ (left), and radial distribution of the individual planet's torque density $T(r)$ and their total  torque (right). The upper two rows show results for  \texttt{fid\_M02\_R1.5} at $t{=}1.5 \times 10^4~P_0$ and $4 \times 10^4~P_0$, and the lower row gives those for \texttt{fid\_M20\_R1.5} at $t{=}1.5 \times 10^4~P_0$. In the right column, the red and blue solid lines are the torque densities of inner and outer planets, whereas the black dashed line represents the total net torque summed over the two planets. The torque profiles are averaged over 20 snapshots within $1~P_0$. In all panels, red and blue circles indicate the positions of the inner and outer planets, respectively. The video can be downloaded from: \protect\url{https://github.com/bbliu-astro/movies/blob/main/hydrodynamic_rebound/fiducial32_p0p2.gif}
    }
    \label{fig:torq_2pL}
\end{figure*}

 \section{Migration of two-planet systems}
\label{sec:pair}

In this section, we present the results of simulations for two-planet pairs, varying both their masses and mass ordering. The simulation parameters are summarized in Table~\ref{tab:planets}. We classify the results into three categories based on the planet masses. In all cases, both planets are inserted into the disk at $t = 5 \times 10^{3}~P_0$ when the inner cavity has just formed and the code is switched to $2$D evolution.   The specific evolution for each pair is detailed in the following subsections.

 \subsection{Low-mass planet pairs}
 \label{sec:low_mass}

\begin{figure*}[t]
    \centering
    \includegraphics[width=0.96\hsize]{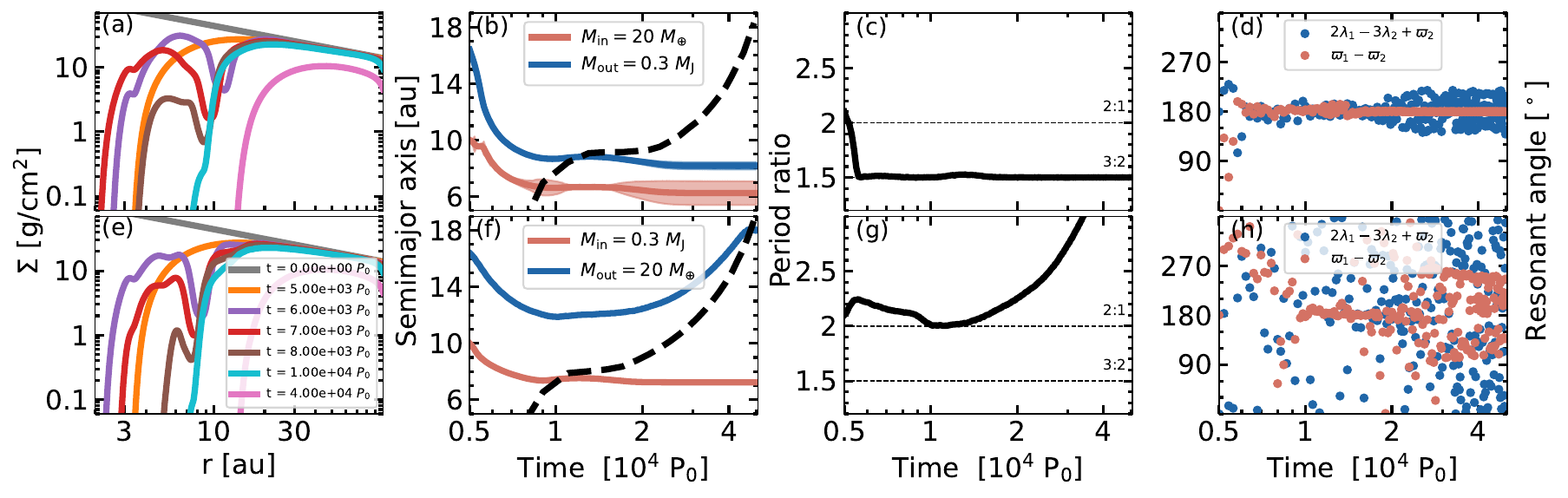}
    \caption{  
Same as Figure 1, but for a two-planet system consisting of $20~\Me$ and $0.3~\Mj$ planets. 
}
    \label{fig:semi_2pM}
\end{figure*}

\begin{figure*}[h!]
    \centering
    \includegraphics[width=0.85\hsize]{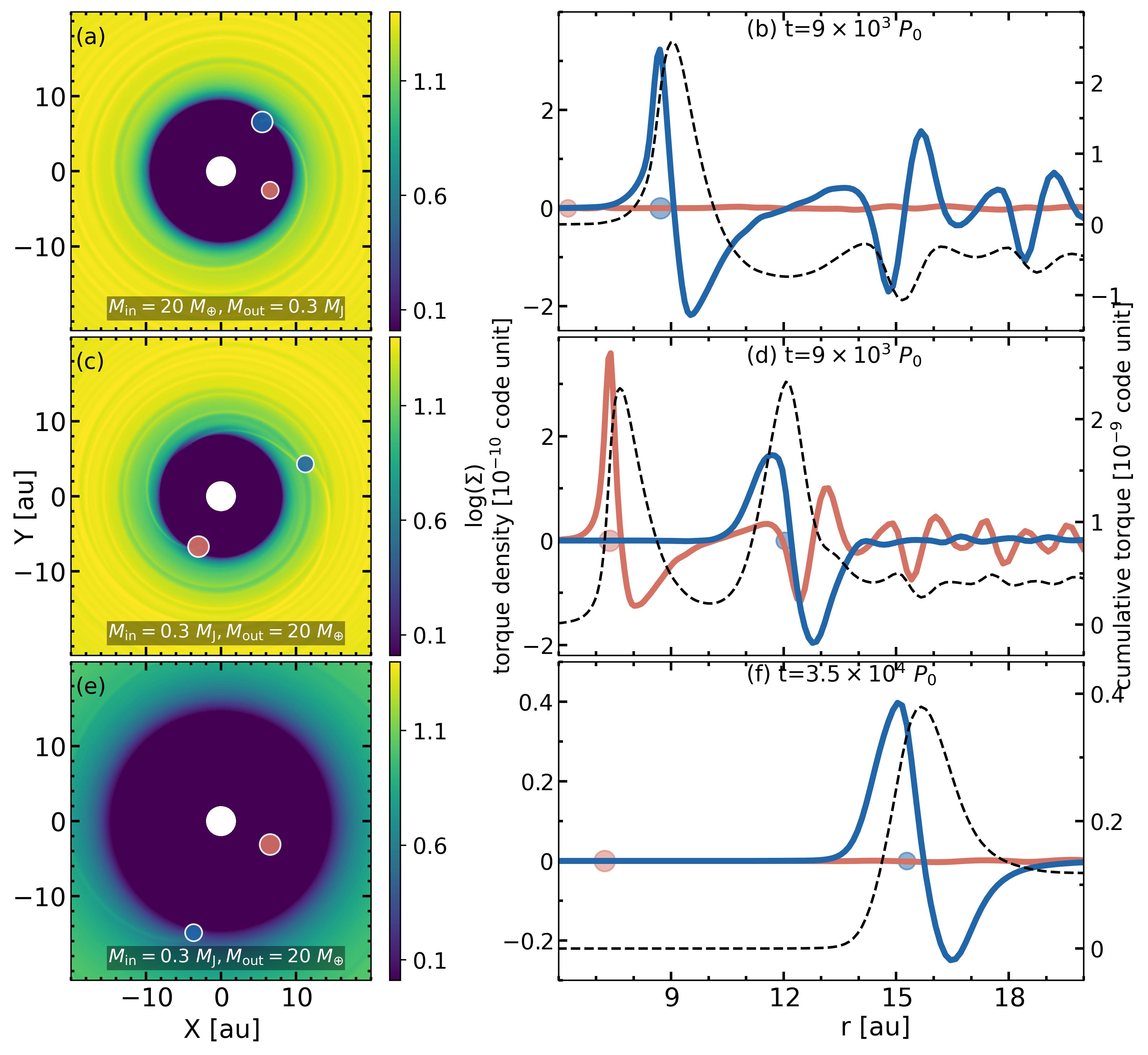}
    \caption{  
 Snapshots from the data in Figure 3 of surface density $\rm \log(\Sigma)$  (left), and radial distribution of the individual planet's torque density $T(r)$ and their total  torque (right). The upper row shows results for \texttt{fid\_M23\_R2.0} at $t{=}9 \times 10^3~P_0$ and the lower two rows give those for  \texttt{fid\_M32\_R2.0} at $9 \times 10^3~P_0$ and $3.5 \times 10^4~P_0$, respectively.}
    \label{fig:torq_2pM}
\end{figure*}

\begin{figure*}[t]
    \centering
    \includegraphics[width=0.96\hsize]{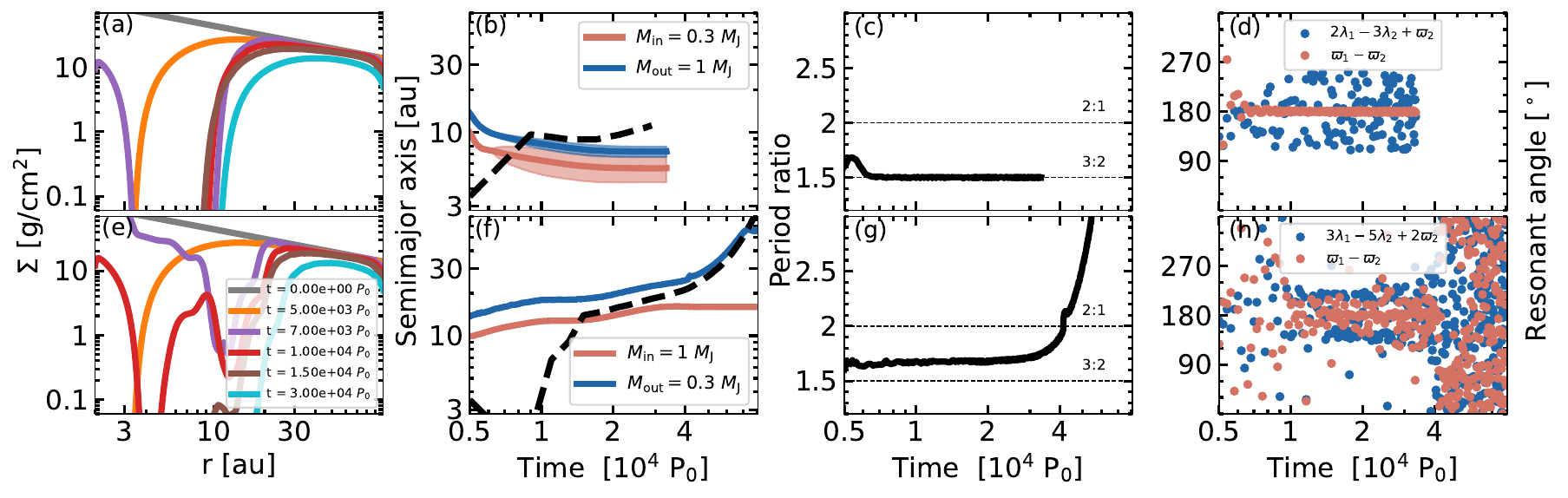}
    \caption{  
Same as Figure 1, but for a two-planet system consisting of $0.3~\Mj$ and $1~\Mj$ planets. 
}
    \label{fig:semi_2pH}
\end{figure*}

\begin{figure*}
    \centering
    \includegraphics[width=0.8\hsize]
    {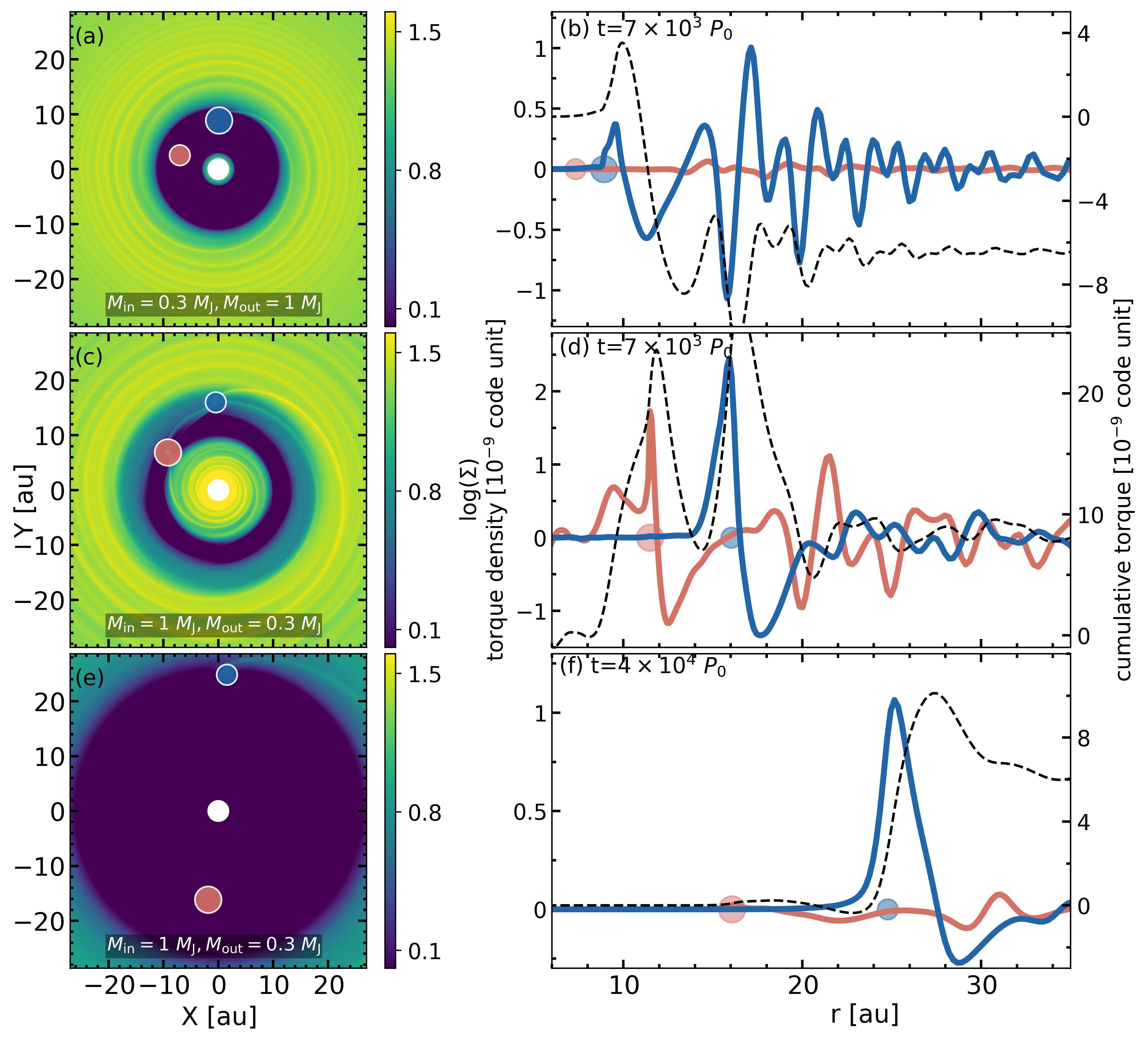}
    \caption{  
  Snapshots from the data in Figure 5 of surface density $\rm \log(\Sigma)$ (left) and radial distribution of the individual planet's torque density $T(r)$ and their total  torque (right). The upper row shows results for \texttt{fid\_M34\_R1.5} at $t{=}7 \times 10^3~P_0$ and the lower two rows give those for \texttt{fid\_M43\_R1.5} at $7 \times 10^3~P_0$ and $4 \times 10^4~P_0$, respectively.}
    \label{fig:torq_2pH}
\end{figure*}

The top panels of Figure~\ref{fig:semi_2pL} illustrate the evolution of the disk surface density and the semimajor axes of planets for a system comprising an inner $5~\Me$ planet and an outer $20~\Me$ planet (\texttt{fid\_M02\_R1.5} in Table~\ref{tab:planets}). Initially positioned with a period ratio slightly above 1.5, the planets migrate inward rapidly and are captured into a 3:2 MMR within a few hundred $P_0$ (Figures~\ref{fig:semi_2pL}b and~\ref{fig:semi_2pL}c). As the inner planet approaches the cavity edge at $t \simeq 1.5 \times 10^{4}~P_0$, its inward migration halts as it enters the low-density cavity.  Subsequently, as the cavity edge sweeps across the outer planet's orbit, that planet reverses its migration direction, moving outward between $t \simeq 2 \times 10^{4}~P_0$ and $4 \times 10^{4}~P_0$. This divergent migration forces the pair out of their initial resonant configuration, as confirmed by the resonant angle of $2\lambda_{1} -3 \lambda_2 + \varpi_2$ ($\lambda$ and $\varpi$ are mean longitude and longitude of periapsis, and the subscripts $1$ and $2$ denote the inner and outer planet) in Figure~\ref{fig:semi_2pL}d, which librates until $t \simeq 2 \times 10^{4}~P_0$ before transitioning to circulation. The outer planet eventually stalls at $16$~au, resulting in a final period ratio of $2.85$.

The bottom panels of Figure~\ref{fig:semi_2pL} show the case with the mass ordering of the two planets reversed (\texttt{fid\_M20\_R1.5}). Here, the less massive outer planet migrates more slowly than its inner counterpart. The inner planet first migrates inward to the cavity and rebounds as the cavity expands. This evolution causes their period ratio to increase within the first $10^{4}~P_0$ before declining afterward. The two planets migrate convergently and are temporarily captured into the 3:2 MMR at $2 \times 10^{4}~P_0$ (Figure~\ref{fig:semi_2pL}g and h). The pair leaves the resonance as the cavity sweeps further outward, causing both planets to sequentially drop into the cavity. However, because the outer planet is less massive, its rebound migration is weaker when the cavity edge sweeps by. Consequently, the planets settle with a final period ratio of $1.7$, only slightly higher than the original resonant value.

Figure \ref{fig:torq_2pL} shows the snapshots of the surface density  $\log(\Sigma)$ (left), and radial profiles of the individual planet's torque density and total torque of the system (right) in \texttt{fid\_M02\_R1.5} and \texttt{fid\_M20\_R1.5}. In the right panels, the red and blue solid lines show the torque density of the inner and outer planets, respectively, while the black dashed line represents their total torque ($\Gamma_{\rm in} + \Gamma_{\rm out}$). The planet pair is close to cavity edge in the top row of Figure \ref{fig:torq_2pL} at $t{=}1.5\times 10^{4}~P_0$. In this case, the outer $20~\Me$ planet features an asymmetric torque density distribution, with the inner peak $10\%$ higher than the outer peak. The torque onto the inner planet is very modest, since it is less massive and very close to the cavity where the density is strongly depleted. Thus, the net torque of the system is positive and driven by the outer planet (Figure~\ref{fig:torq_2pL}b). This divergent migration causes the planet pair to move out of the resonance (Figure \ref{fig:semi_2pL}b). As the cavity further retreats, the inner planet does not migrate outward fast enough, leaving only the outer planet rebound. 
This can also be seen in the middle row of Figure \ref{fig:torq_2pL}. The outer planet experiences a stronger, asymmetric positive torque as the cavity approaches it at $t{=}4\times 10^{4}~P_0$.

On the other hand, in the case with an inner massive planet at $t{=}1.5\times 10^{4}~P_0$ (bottom row of Figure \ref{fig:torq_2pL}), the inner massive planet located at the disk edge experiences a much stronger positive torque. For the outer low-mass planet, its torque contributions from the inner and outer peaks nearly cancel out. The total torque for the system is positive. In this circumstance the migration is convergent, which differs from that in the top row of Figure~\ref{fig:torq_2pL}.

It is important to note that this work adopts a broad definition of rebound migration, encompassing any scenario where a planet migrates outward in tandem with an expanding disk cavity. Previous studies have shown that for low-mass planets, this outward migration is dominated by the corotation torque, driven either by a positive density gradient near the boundary \citep{Masset2006}, or a sharp boundary transition when only upper U-turn is present within the planetary horseshoe region \citep{Liu2017a}. 
In the former case, the corotation torque is two-sided and driven by a positive vortensity gradient, whereas in the latter limit, the angular momentum exchange in the corotation region comes solely from the upper horseshoe. Alternatively, for gap-opening planets, the required positive torque can originate from asymmetric Lindblad torques induced by disk eccentricity \citep{Dempsey2021,Liu2025}. The aim of this work is not to specify which underlying mechanism drives outward migration in each specific case, but highlight this resonance breaking channel through rebound.

 \subsection{Low-mass and massive planet pairs}
 \label{sec:mixed_mass}

We also performed simulations with a low-mass planet of $20~\Me$ and a giant planet of $0.3~\Mj$. The top panels of Figure~\ref{fig:semi_2pM} show the results from \texttt{fid\_M23\_R2.0}, where the inner planet mass is significantly lower than that of the outer planet. Although the pair starts with a period ratio slightly beyond 2, the outer massive planet opens a partial gap and undergoes rapid type III migration. Consequently, the planets easily bypass the $2$:$1$ MMR and become trapped in the $3$:$2$ MMR. Both planets migrate inward and fall into the disk cavity at $t {=} 9\times 10^{3}~P_0$ and $1.1\times10^{4}~P_0$, respectively.

Their resonant state survives through the inside-out disk clearing phase. Notably, as the disk dissipates, the resonant state transitions from stable to overstable \citep{Goldreich2014,Lin2025}. This is evidenced in Figure~\ref{fig:semi_2pM}d by the resonant angle switching from a low amplitude libration around $180^{\circ}$ to an increased amplitude of $60^{\circ}$, accompanied by an elevated eccentricity of the inner super-Earth planet from $<0.01$ to $0.12$ (Figure~\ref{fig:semi_2pM}b).

We observe an entirely different evolutionary pattern when the mass ordering of the planet pair is reversed (bottom panels of Figure~\ref{fig:semi_2pM}). The planets initially migrate divergently until the inner giant planet enters the cavity. The outer super-Earth continues inward migration and is temporarily trapped in a 2:1 resonance. However, as the cavity edge sweeps outward, it pushes the super-Earth planet away, leaving the final system substantially displaced from the resonance.

The torque density and cumulative distributions are further analyzed in Figure~\ref{fig:torq_2pM}. In the configuration with a less massive inner planet (top row of Figure~\ref{fig:torq_2pM}), at $t{=}9\times 10^{3}~P_0$ the inner planet lies within the cavity, whereas the outer planet opens a partial gap that overlaps with the gas clearing driven by photo-evaporation. Under these conditions, the outer planet experiences a overall negative torque, causing both planets to eventually migrate into the inner disk cavity.

Conversely, in the case with a more massive inner planet at $t{=}9 \times 10^{3} \ P_0$, the inner planet experiences a net positive torque at the disk edge, while the outer planet experiences a nearly zero net torque farther from the edge (middle row of Figure~\ref{fig:torq_2pM}).  As can be seen from Figure~\ref{fig:semi_2pM}e and f, the cavity quickly sweeps through the inner planet's orbit from $t{=}8 \times 10^{3} \ P_0$ to $10^{4} \ P_0$. Although the total torque for the pair is positive, the migration speed induced by the net torque in this case is lower than the cavity expansion rate. Consequently, the inner planet cannot keep pace with the outward-moving cavity and falls into it.
At later times the disk only carries the outer, less massive planet, which continues to surf the expanding cavity until $t{\sim}4\times 10^{4}~P_0$.   For example, at $t{=}3.5\times 10^{4}~P_0$, the positive torque peak on the outer planet is nearly twice as strong as the negative torque peak (bottom row of Figure~\ref{fig:torq_2pM}).

 \subsection{Massive planet pairs}
 \label{sec:massive}

In the simulations with an inner planet of $0.3~\Mj$ and an outer planet of $1~\Mj$, the pair is quickly captured into the 3:2 MMR. Their period ratio remains locked at $1.5$ while their eccentricities are excited to ${\sim}0.1{-}0.2$ during disk dispersal (Figure~\ref{fig:semi_2pH}b and d). The pair remains in the resonant state but with large-amplitude libration of the resonant angles, which is a signature of overstability \citep{Lin2025}. This fragile nature, associated with large-amplitude libration and elevated eccentricities, indicates that the system might be long-term unstable \citep{Hu2025}. However, as shown in Figure~\ref{fig:semi_2pH}f, when the two planet masses are reversed, both planets undergo outward migration from the beginning. They reside in the 5:3 MMR until $t \sim 2.5 \times 10^{4}~P_0$. As the disk edge continues to retreat, the inner Jupiter-mass planet enters the cavity, leaving only the outer Saturn-mass planet to migrate outward and further increasing their period ratio (Figure~\ref{fig:semi_2pH}g).

\begin{figure}
    \centering
       \includegraphics[width=0.96\hsize]{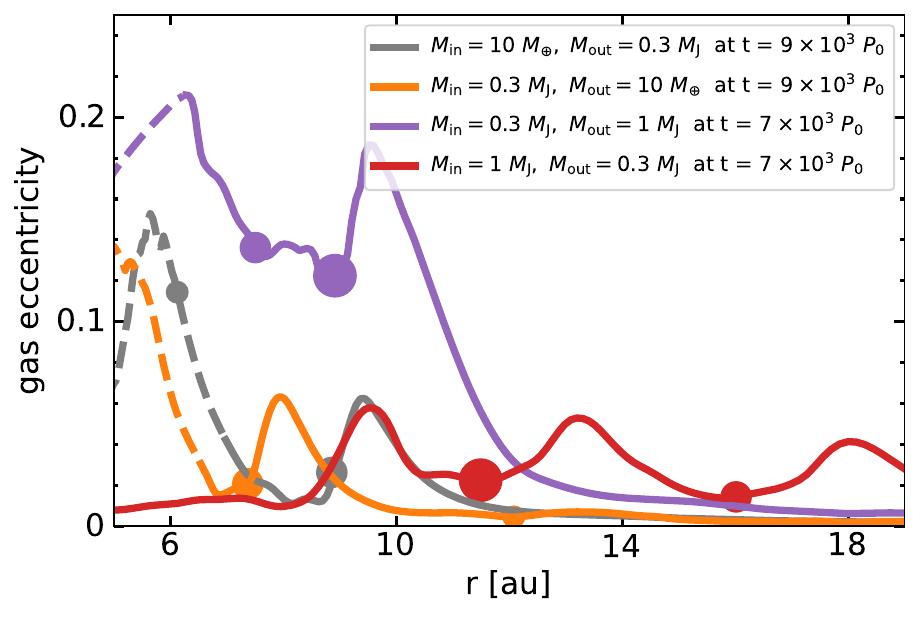}   
    \caption{Azimuthally averaged radial distribution of the disk eccentricity. The filled circles show the planet's location in each simulation. The dashed part of the curves indicate the gas-depleted cavity region, where the calculation of the gas eccentricity is strongly affected by the reset in the gas density. The Saturn-Jupiter planet pair excites a large gas eccentricity, in contrast to the other three cases.} 
    \label{fig:ecc}
\end{figure}

In contrast to the low-mass cases discussed previously, the presence of two massive planets induces strong perturbations in the disk gas, as can be seen in Figure~\ref{fig:torq_2pH}.
Following the method of paper I, we compute the azimuthally averaged radial eccentricity profile of the disk. Treating the gas as pressure-less particles on Keplerian orbits around the central star\footnote{ In reality, the gas flow is slightly sub-Keplerian due to the pressure gradient. The pressure-less particle approximation may introduce an artificial eccentricity, see \cite{Teyssandier2017}. }, we first calculate the eccentricity for each gas cell from its local position and velocity. The eccentricity of the gas ring at radius $r$ is then obtained by azimuthally averaging these individual cell values.
Figure~\ref{fig:ecc} illustrates the disk eccentricity for various planet pair configurations. Notably, the disk becomes significantly eccentric with $e_{\rm g}{\sim}0.15$ for a Saturn-Jupiter pair in \texttt{fid\_M34\_R1.5} (purple dots in Figure~\ref{fig:ecc}). Consequently, the resulting torque density profile departs from the classical two-lobed peak structure, instead exhibiting oscillations driven by large-amplitude non-Keplerian gas motions (top row of Figure~\ref{fig:torq_2pH}). At $t {=} 7 \times 10^{3}~P_0$, the torque originates primarily from the outer disk, yielding a net negative Lindblad torque that drives both planets inward.

When the Jupiter-Saturn pair are in nearly circular orbits (red dots in Figure~\ref{fig:ecc}), the principal term contributes to the disk eccentricity evolution (e.g., see Eq. 13 of \citealt{Teyssandier2016} with $l{=}m$). For massive planets, gas in the horseshoe region is significantly depleted, eliminating effects from the corotation resonance. Generally, the outer Lindblad resonance excites eccentricity, while the inner Lindblad resonances damp it down \citep{Goldreich2003,Masset2008,Teyssandier2016}.  Since the inner disk is quickly removed through photoevaporation, only the outer disk remains, which causes disk eccentricity to increase (purple line in Figures~\ref{fig:ecc}).

When the inner planet is more massive, however, the inner disk is not fully depleted before $t{=} 10^{4}~P_0$ (Figures~\ref{fig:semi_2pH}e) and the disk still remains largely circular (red line in Figures~\ref{fig:ecc}).
The two planets open a common gap, and the torques onto the two planets within the gap region nearly cancel out. The net torque on the system is thus mainly determined by the disk gas on both sides of the planets. Because the inner planet is more massive than the outer one, the inner residual disk exerts a stronger positive integrated torque onto the Jupiter-mass planet that overcomes the negative torque from the outer disk onto the Saturn-mass planet (see black dashed line in Fig.~\ref{fig:torq_2pH}d). As a result, the system experiences a net positive torque, driving the outward migration of both planets.
This migration pattern was first identified by \cite{Masset2001}. We confirm that the similar physics also operates in the photoevaporating disk with an expanding cavity.  These two planets migrate outward together up to $t {=} 4 \times 10^{4}~P_0$. After that, the Jupiter-mass planet falls into the disk cavity, while the Saturn-mass planet retains a net positive torque (bottom row of Fig.~\ref{fig:torq_2pH}).

 \section{Migration of multi-planets}
\label{sec:multi}

Here we discuss the simulations for three-planet systems. The masses of the planets from the innermost to the outermost are $10~\Me$, $20~\Me$ and, $0.3~\Mj$. We explore their migration with two gas disk configuration: a fiducial model ($\Sigma_0{=}44 \rm \ g/cm^2$) and a model with lower gas density ($\Sigma_0{=}13 \rm \ g/cm^2$). Initially, all neighboring planet pairs are placed close to their $3$:$2$ MMRs. 

\begin{figure*}[t]
    \centering
    \includegraphics[width=0.96\hsize]{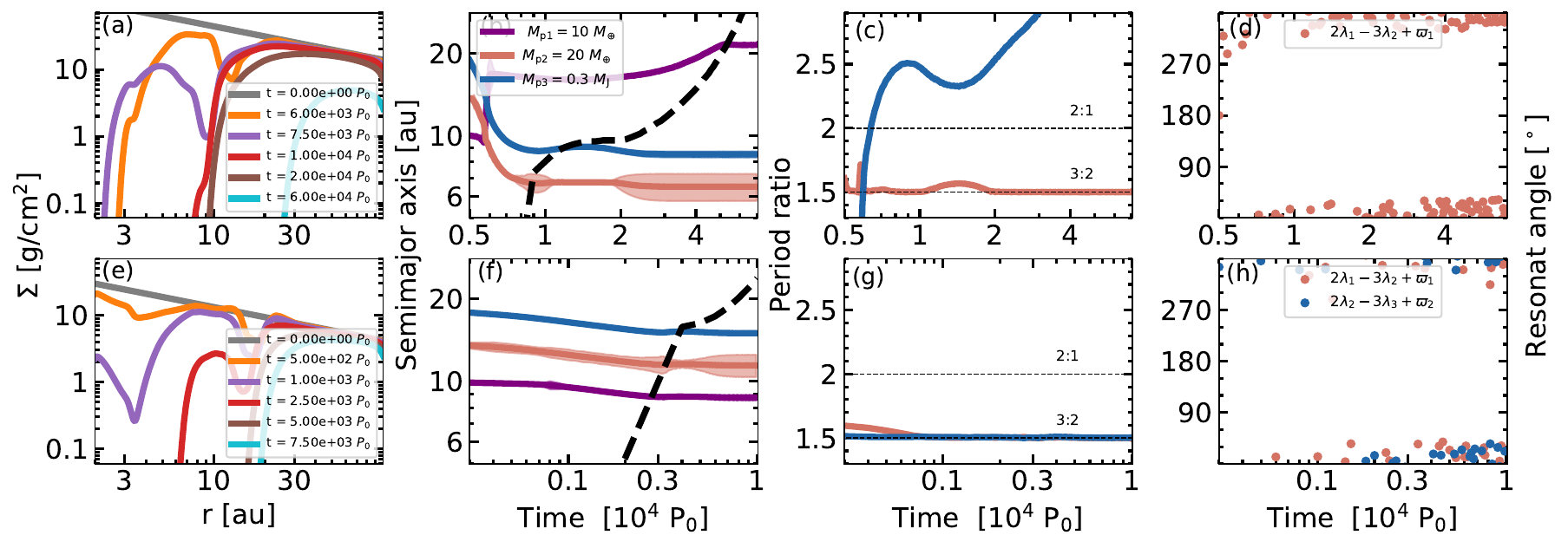}
    \caption{  
Evolution of disk surface density, planet semimajor axes, period ratios, and resonant angles for a three-planet system consisting of $10~\Me$, $20~\Me$ and $0.3~\Mj$ planets. The top and bottom rows show the configurations of the fiducial disk model and less massive disk model.
In the second column, the semimajor axes of the planets from the inner to outer are given by the red, blue and purple curves, respectively, with shaded regions representing their perihelion and aphelion distances. The black dashed line marks the inner disk edge, defined as the radius where the surface density falls below $0.1~\rm g/cm^2$. In the third column, the blue and red lines give the outer-to-inner planet period ratio for the inner and outer pairs, respectively. The fourth column plots the resonant angles of the inner (red) and outer (blue) planet pairs.
}
    \label{fig:comparison}
\end{figure*}

The top panels of Figure~\ref{fig:comparison} illustrate the evolution in the fiducial disk model. The outer two planets are captured into a  $3$:$2$ resonance and migrate rapidly inward. This leads to a close encounter that scatters the innermost planet into the outer disk region even in the gas-rich disk phase (also see \citealt{Marzari2010}). The inner two planets of  $20~\Me$ and $0.3~\Mj$ stay locked in the resonance as the disk cavity retreats outward. When the cavity approaches the outermost planet, it rebounds farther out. As seen in this proof-of-concept simulation, the orbital architecture is altered during disk dispersal, and the system evolves from initial compact configuration into final one consisting of a resonant pair and an isolated low-mass planet with wider orbital separations.

The bottom panels of Figure~\ref{fig:comparison} show results for the same planet configuration but in a lower-mass disk. In this case, disk dispersal proceeds much faster. Convergent migration results in planets sequential trapping into  $3$:$2$ resonances. Because the cavity expands too rapidly, no rebound occurs during gas dissipation, leaving the resonant chain intact. Combined with the analysis of Paper I, these results demonstrate once again that rebound migration requires specific conditions. It does not occur if the cavity expands either too quickly or too slowly, the rate of which depends on the gas disk mass and/or the strength of stellar photoevaporation (see Figure 7 of Paper I).

\begin{figure*}[t]
    \centering
        \includegraphics[height=0.5\textwidth]{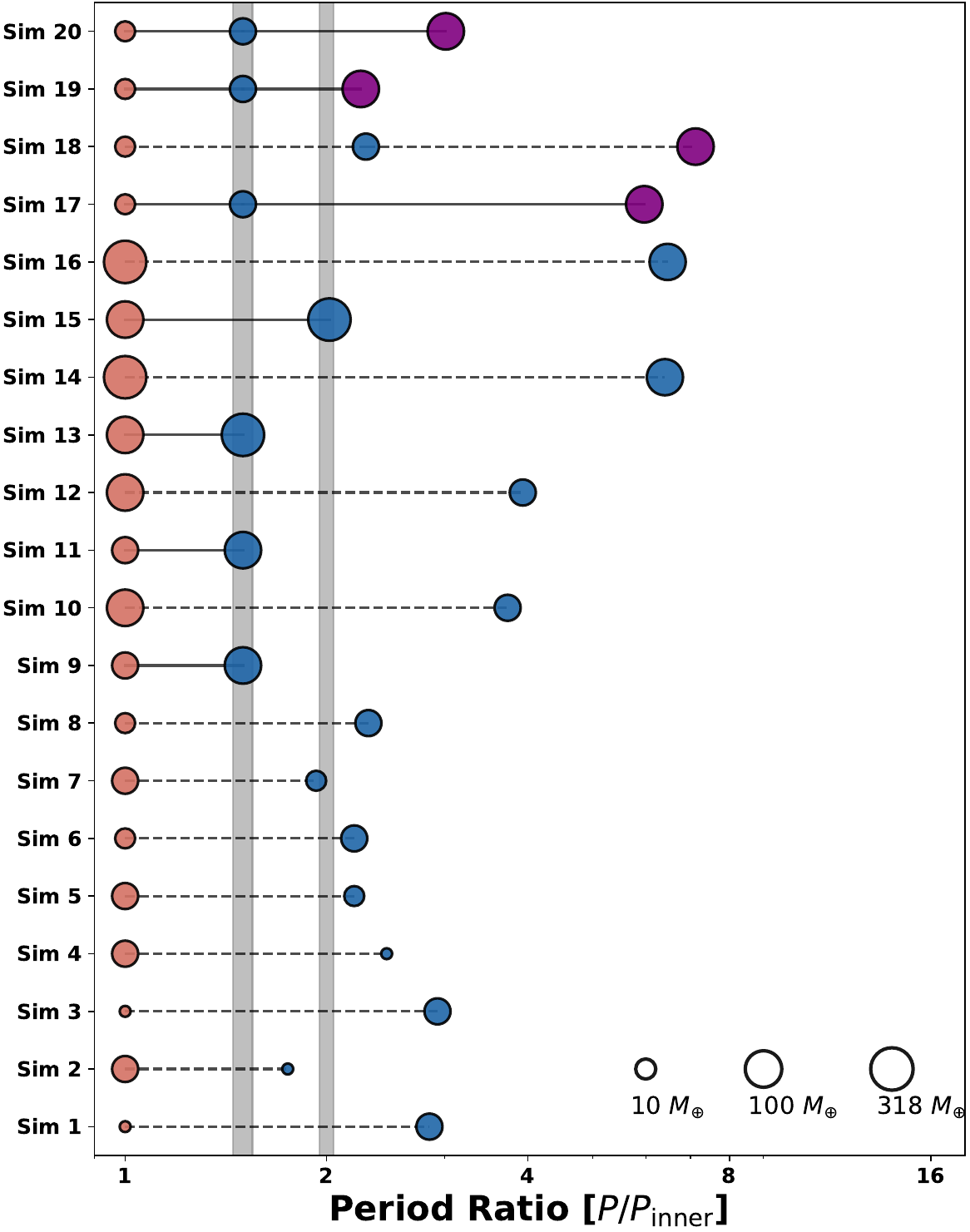}
    \includegraphics[height=0.5\textwidth]{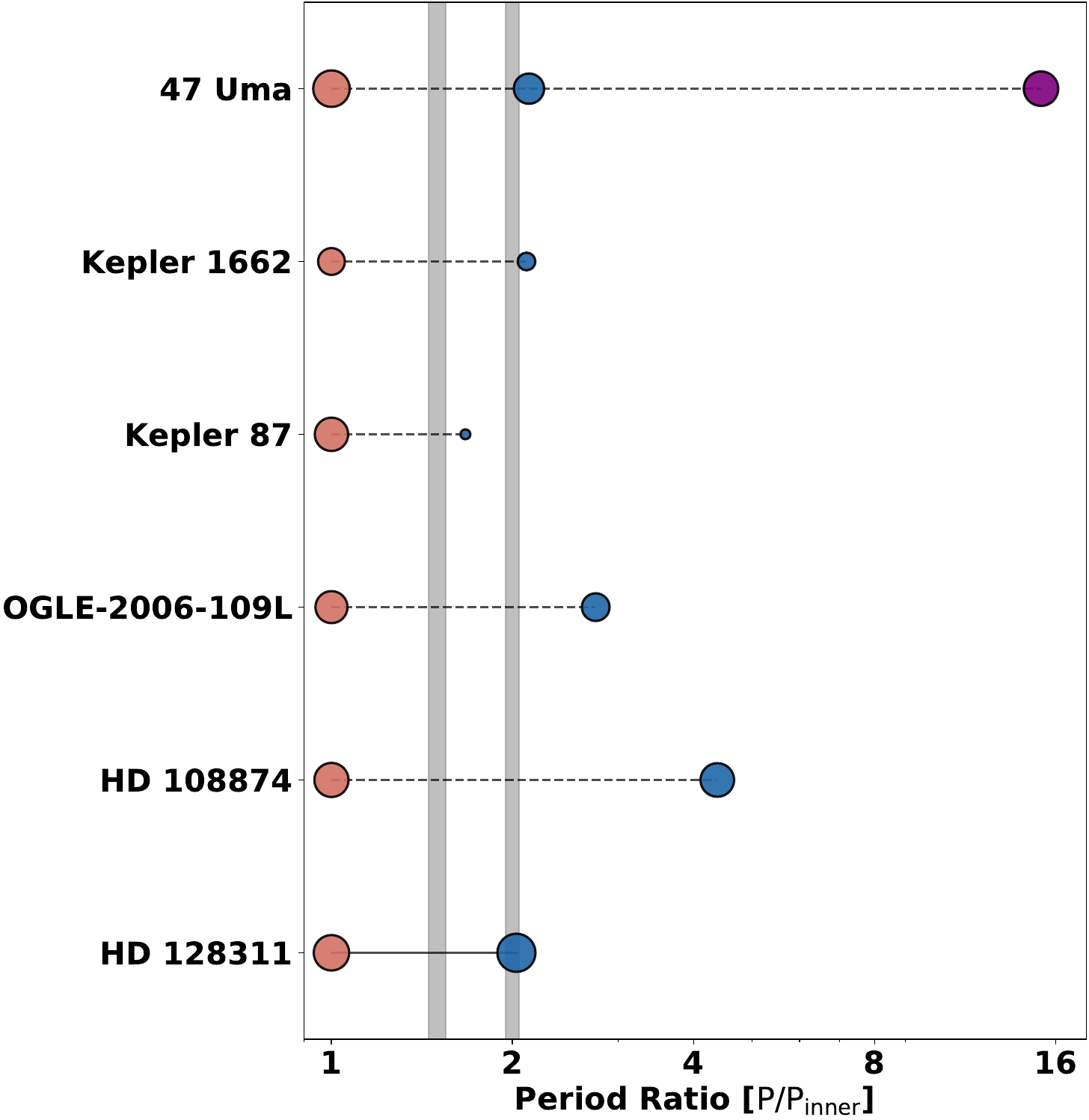}
    \caption{ Period ratios of all simulated systems (left) and a selection of observed extra-solar systems (right). Planets are ordered from inner to outer, color-coded as orange, blue, and purple, respectively, with symbol size representing planet mass. The solid horizontal line marks systems that remain in resonance, while the dashed line indicates period ratios that deviate from resonance which may be driven by rebound migration. The vertical gray shaded regions denote the $3$:$2$ and $2$:$1$ mean-motion resonances. The observed systems are around sun-like stars, with the innermost planet having an orbital period longer than $100$ days. 
}
    \label{fig:simobs}
\end{figure*}

Now we summarize all $20$ simulations in Table~\ref{tab:planets} for systems with both two or three planets.  The left panel of Figure~\ref{fig:simobs} presents simulated systems with their masses and period ratios, while the right panel shows a selection of observed exoplanet systems for comparison. Since our simulations focus on solar-mass stellar hosts and the photoevaporation mainly alter the planets  whose orbits are on the order of 1-10 au,  we select the observed systems around solar-like stars where the innermost planet has an orbital period longer than $100$ days. From inner to outer, planets are color-coded orange, blue, and purple, respectively, with symbol size corresponding to planet mass.
The solid horizontal line marks systems that remain in resonance, whereas the dashed line indicates period ratios that have deviated from resonance. Overall, these cases demonstrate the potential of rebound migration to reshape system architectures. Further studies are warranted to explore in depth how the period ratios of systems depend on planet masses and their ordering.

\section{Discussion}
\label{sec:discussion}

Our simulations assume a locally isothermal equation of state, meaning the disk temperature varies only with radial distance and remains constant in time. Under this approximation, the corotation torque exerted on the planets is driven entirely by the vortensity gradient. 
While this approximation is widely used in planet–disk interaction studies, it neglects the thermal evolution of the gas and the entropy‑related effects that can contribute to the corotation torque \citep{Paardekooper2010,Paardekooper2011}.
In reality, protoplanetary disks possess finite cooling timescales. The inclusion of a full energy equation would introduce entropy-related corotation torques, which could modify the net torque balance and alter the planets' migration rates.

The current study relies solely on an X-ray photoevaporation prescription, adopting the mass-loss rate profile from \cite{Owen2012}. However, alternative dispersal mechanisms such as far-ultraviolet (FUV) photoevaporation \citep{Gorti2015} or magnetically driven disk winds \citep{Bai2016,Suzuki2016} could produce fundamentally different cavity structures, edge gradients, and overall dispersal timescales \citep{Pascucci2023}. Because the efficacy of rebound migration is highly sensitive to the speed at which the cavity expands, a systematic study of the rebound efficiency under various dispersal models is an important direction for future work.

Furthermore, throughout our 2D simulations, planets are introduced at their full masses without gas accretion. As recently demonstrated by \cite{Ida2026}, gas accretion onto a planet can significantly alter the surrounding gas flow and its resulting migration torque. For our massive planet cases in particular, incorporating realistic gas accretion rates could further deplete the local gas surface density, potentially shifting the delicate balance between Lindblad and corotation torques during the critical rebound phase. 
Although simulations that self-consistently include gas accretion onto planets are beyond the scope of this paper, such models are needed to fully capture the migration physics of massive planets.

Finally, for computational feasibility, we restricted our simulations to a fixed turbulent viscosity parameter of $\alpha {=} 10^{-3}$ and disk aspect ratio of $h_0 {=} 0.05$. The rebound mechanism, as demonstrated in Paper I, is known to depend on these parameters. Moreover, we have only considered two initial period ratios ($3$:$2$ and $2$:$1$ resonances) and a limited set of planet mass combinations. A more comprehensive parameter study is beyond the computational scope of the present work, but we caution that our results may not be fully representative of the entire diversity of protoplanetary disks and planetary systems. In particular, the sensitivity of rebound migration to the disk dispersal timescale (illustrated by the comparison between the fiducial and low‑mass disk models in Section 4) suggests that a careful exploration of the cavity expansion rate, which depends on both the photoevaporation strength and the initial disk mass, is essential for robust statistical predictions. 

\section{Conclusions}
\label{sec:conclusion}

In this work, we have extended our 2D hydrodynamical investigation of disk-planet interactions to explore how rebound migration operates in multi-planet systems near the expanding cavity of a photoevaporating disk. Using the Dusty FARGO-ADSG code with a source term in the continuity equation that mimics the mass loss rate obtained from \cite{Owen2011,Owen2012}, we simulated systems of two and three planets spanning a wide mass range, from super-Earths to Jupiters, under varying disk conditions.

Our simulations confirm that rebound migration, driven by the strong positive corotation torque at the cavity edge, remains effective in multi-planet environments, where it can significantly reshape system architectures. The outcome depends critically on planet masses, mass ordering and disk parameters.

\begin{itemize}
    \item In low-mass pairs (e.g., $5$–$20~M_\oplus$), divergent migration driven by the expanding cavity can pull planets out of resonance, leading to period ratios significantly larger than the initial resonant values. Rebound is most pronounced for the outer planet when it is more massive (Figures 1 \& 2).
    \item In mixed-mass pairs (e.g., $20~M_\oplus$ and $0.3~M_{\rm J}$), rapid migration of a massive outer planet can bypass wider resonances and establish a stable $3:2$ MMR that survives disk clearing. When the inner planet is more massive, however, divergent migration caused by cavity expansion can leave the system displaced from resonance (Figures 3 \& 4).
    \item In massive pairs (e.g., $0.3~M_{\rm J}$ and $1~M_{\rm J}$), strong disk perturbations and eccentricity excitation can modify torque profiles, yet outward migration is still possible when the two planets open a common gap and the inner planet is sufficiently massive to maintain a net positive torque (Figures 5, 6 \& 7).
\end{itemize}

In three-planet systems, we find that rebound migration can alter the system's architecture during disk dispersal. In our fiducial disk, due to rapid inward migration of the outer two massive planets, the inner planet undergoes orbital crossing and later gets scattered outward. The outermost planet later rebounds as the cavity sweeps past it, yielding a final architecture of a close-in resonant pair and a distant, lower-mass planet. In contrast, in a lower-mass disk where photoevaporation drives faster cavity expansion, rebound is suppressed. Here, convergent migration results in the sequential capture and maintenance of a resonant chain without instability, highlighting the sensitivity of system outcomes to disk dispersal timescales ((Figure 8).

 This study provides the first hydrodynamical evidence that rebound migration can critically influence the orbital evolution and final architectures of multi-planet systems during the late stages of disk dispersal. By breaking resonances and widening period ratios through divergent migration, rebound migration offers a robust mechanism to transition systems from tightly packed, primordial resonant chains into the non-resonant configurations that characterize much of the observed exoplanet population ((Figures 9). Therefore, the late-stage disk dispersal phase is a critical epoch for defining the final observational signatures of multi-planet systems. While a detailed statistical comparison with observational data is beyond the scope of this study, future works should further explore the parameter space of planet masses, disk properties, and photoevaporation models, as well as the long-term stability of systems shaped by rebound migration.

\begin{acknowledgements} 
We thank the referee's careful reading and helpful suggestions to improve the manuscript. This work is supported by the National Key R\&D Program of China (2024YFA1611803). BL acknowledges support from the National Natural Science Foundation of China (Nos. 12222303 and 12173035) and the start-up grant of the Bairen program from Zhejiang University. The simulations and analysis presented in this article were carried out on the SilkRiver Supercomputer of Zhejiang University.
Y.-P. L. is supported in part by the Natural Science Foundation of China (Grants No. 12373070 and No. 12192223), and the Natural Science Foundation of Shanghai (Grant No. 23ZR1473700).
\end{acknowledgements}

\bibliographystyle{aa}
\bibliography{reference.bib}
\end{document}